\RequirePackage{fix-cm}
\RequirePackage{bbm}
\RequirePackage{float}
\RequirePackage{physics}
\RequirePackage{amssymb}
\RequirePackage{bbold}
\RequirePackage{bm}
\RequirePackage[table]{xcolor}
\RequirePackage{comment}
\RequirePackage{siunitx}
\RequirePackage{multirow}
\RequirePackage{hhline}
\RequirePackage{lineno}
\documentclass[smallcondensed]{svjour3}     

\smartqed  
\RequirePackage{graphicx}
\RequirePackage[colorlinks=true, linkcolor=black, urlcolor=blue, citecolor=gray]{hyperref}
\journalname{Hyperfine Interactions}

\newcommand{\ecm}{\textit{e$\cdot$cm}}

\begin{document}

\title{A survey of nuclear quadrupole deformation in order to estimate the nuclear MQM and its relative contribution to the atomic EDM}
\subtitle{Searching for the best candidate to focus on an atomic EDM measurement: A nuclear physics perspective}

\titlerunning{Survey of nuclear quadrupole deformation to estimate atomic EDM}

\author{
Prajwal MohanMurthy$^{1,*}$ \and
Umesh Silwal$^{2}$ \and
Jeff A. Winger$^{3,\dagger}$
}


\institute{
$^1$ Department of Physics, University of Chicago, Chicago, IL 60637, U.S.A\\
$^*$ Corresponding author, \email{prajwal@alum.mit.edu}, Present address: Laboratory for Nuclear Physics, Massachusetts Institute of Technology, 77 Mass. Ave., Cambridge, MA 02139, USA\\
$^2$ Department of Physics and Optical Science, University of North Carolina, Charlotte, NC 28223, USA\\
$^3$ Dept. of Physics and Astronomy, P.O. Box 5167, Mississippi State University, Mississippi State, MS 39762, USA\\
$^{\dagger}$ Corresponding author, \email{j.a.winger@msstate.edu}
}

\date{Received: [Ver5 - \today] / Accepted: -}

\maketitle

\begin{abstract}
New sources of charge-parity (CP) violation, beyond the known sources in the standard model (SM), are required to explain the baryon asymmetry of the universe. Measurement of a non-zero permanent electric dipole moment (EDM) in fundamental particles, such as in an electron or a neutron, or in nuclei or atoms, can help us gain a handle on the sources of CP violation, both in the SM and beyond. 
The nuclear magnetic quadrupole moment (MQM), the central topic of this work, is also CP, P, and T violating. Nucleons and nuclei have a non-zero MQM from sources within the SM, but the nuclear MQM is dramatically enhanced if the nuclei are structurally quadrupole deformed. Multiple sources contribute to an atomic EDM \emph{viz.} (i) nuclear EDM through its Schiff moment, which is enhanced by nuclear octupole deformation, (ii) CP violating interactions between the electrons and the nuclei, and (iii) the nuclear MQM that contributes to the atomic EDM in atoms with an unpaired valence electron. 
Our survey of nuclear quadrupole deformation has identified $^{151}$Nd, $^{153}$Pm, $^{153}$Sm, $^{157}$Ho, $^{163,165}$Er, $^{161,168}$Tm, $^{167}$Yb, $^{169}$Hf, $^{171,180}$Ta, $^{173,175,177,179,180}$Re, $^{190,192}$Ir, $^{188}$Au, $^{223,225}$Fr, $^{223,227,229,231}$Ra, $^{223,225,227,229}$Ac, $^{229,231}$Th, $^{229,231,233,235}$Pa, $^{235}$U, $^{233,235,237,238,239}$Np, $^{237}$Pu, and $^{239,241,242,243,245}$Am as ideal systems in which to search for a CP violating EDM via their enhanced nuclear MQM, while $^{223,225}$Fr, $^{223}$Ra, $^{223,225,227}$Ac, $^{229}$Th, and $^{229}$Pa also have maximally enhanced nuclear Schiff moment contribution due to their octupole deformation. Laser cooling of the isotopes of Er, Tm, Yb, Fr, and Ra has already been demonstrated, making $^{223,225}$Fr and $^{223}$Ra some of the best systems in which to measure an EDM.

\end{abstract}

\section{Introduction}


We exist! According to the three Sakharov conditions \cite{Sakharov1967-vt}, the thus required baryon asymmetry of the universe demands: (i) baryon number violation, (ii) C and CP violation, and (iii) underlying interactions to be out of the thermal equilibrium. Remnant baryon and photon densities of the universe are remarkably well measured using the cosmic microwave background \cite{Aghanim2020-cm}. But the measured baryon asymmetry of the universe is nearly $8$ orders of magnitude larger than that predicted by the Standard Model using sources of CP violation in the CKM-matrix \cite{Riotto1999-za}. This demands additional sources for baryonic CP violation.

Sub-atomic particles, like electrons and neutrons, as well as atoms can have large magnetic dipole moments \cite{Tiesinga2021-qm}. Measurement of a permanent non-zero electric dipole moment (EDM) in subatomic particles and atoms, in addition to their non-zero magnetic dipole moment (MDM), is an indication of P, T, and CP violation. Table~\ref{tab1} indicates the effect of the discrete transforms of C, P, and T upon some relevant vector parameters of electric and magnetic fields, as well as electric and magnetic dipole moments. Even though the MDM is C and T odd, when taken in combination with the fact that the magnetic field is also C and T odd, it does not violate P, CP, or T. But EDM is not simultaneously odd and even \emph{w.r.t.} the magnetic field, indicating P, CP, or T violation. Therefore, it is important to note that a permanent non-zero EDM in a particularly non-scalar system, one whose spin is non-zero, is key as a signature of P, CP, and T violation.

\begin{table}[b]
\centering
\caption[]{Effect of C, P, and T transforms on vector parameters of electric and magnetic fields and dipole moments.}
\label{tab1}
\begin{tabular}{l c c c c c c}
\hline
Vector parameter & $\hat{P}$ & $\hat{C}$ & $\hat{T}$ \\
\hline
Electric Field ($\vec{E}$) & $-$ & $-$ & $+$ \\
Magnetic Field ($\vec{B}$) & $+$ & $-$ & $-$ \\
Electric dipole moment ($\vec{d}$) & $-$ & $-$ & $+$\\
Magnetic dipole moment ($\vec{\mu}$) & $+$ & $-$ & $-$ \\
\hline
\end{tabular}
\end{table}

The only source of baryonic CP violation in the SM is within the CKM-matrix, which describes the weak interactions \cite{Cabibbo1963-kf,Kobayashi1973-tx}. This source of CP violation can be used to estimate the permanent EDM of neutrons and electrons: $d^{\text{(CKM)}}_n\approx2\times10^{-32}~$\ecm~\cite{Khriplovich1982-jh} and $d^{\text{(CKM)}}_e\approx5.8\times10^{-40}~$\ecm~\cite{Yamaguchi2020-on}, respectively. No non-zero EDM of such particles has been measured yet. The current experimental upper bound on neutron and electron EDMs (at 90\% C.L.) are $d_n\!<\!1.8\times10^{-26}~$\ecm~\cite{Abel2020-jr} and $d_e\!<\!4.1\times10^{-30}~$\ecm~\cite{Roussy2023-vd}. However, strong interactions could also violate CP, via the Quantum-Chromo-Dynamics (QCD) $\bar{\theta}$ parameter \cite{t_Hooft1976-lq}. There could also be mechanisms beyond the SM, like super-symmetry, that could add to these sources of CP violation \cite{Ellis1989-sp}. Therefore, measuring a non-zero permanent EDM in subatomic particles or atoms, helps pin down the sources of CP violation, and aid in the search for additional sources of baryonic CP violation beyond the CKM-matrix. The estimates of EDMs of various systems within the SM, like that of neutrons, electrons and atoms, using sources of CP violation in the CKM-matrix and QCD-$\bar{\theta}$, have been well documented \cite{Mohanmurthy2021-ou}.

Atomic EDMs could arise from the EDM of nucleons and electrons, as well as CP violating interactions within the nucleus and between the electrons and nucleus \cite{Chupp2019-cm}. An atomic EDM can also be generated by other CP violating electromagnetic moments of the nucleus, \emph{viz.} the magnetic quadrupole moment (MQM), being the lowest order such magnetic moment that could be non-zero \cite{Sushkov1984-cb}. The relative atomic EDM arising from the contribution of an enhanced nuclear Schiff moment, due to octupole and quadrupole deformations of the nucleus, was previous surveyed and the isotopes of $^{221}$Rn, $^{221,223,225,227}$Fr, $^{221,223,225}$Ra, $^{223,225,227}$Ac, $^{229}$Th, and $^{229}$Pa were identified as attractive candidates \cite{Mohanmurthy2020-np}. In this work, we have surveyed the quadrupole deformation of nuclei to identify candidate isotopes  whose atomic EDM arising from the contribution of the nuclear MQM is maximally enhanced.
Many of these exotic and highly deformed isotopes can, for the first time, be produced in large numbers at the Facility for Rare Isotope Beams (FRIB). Furthermore, certain isotopes lend themselves to laser based cooling, owing to their atomic structure. Cooling can help increase the number density, and ergo the measurable sensitivity to an atomic EDM. We present here new candidate isotopes, in light of the rates deliverable at FRIB, as well as laser cooling considerations.

\section{Multipole Expansion of the Nuclear Potentials}

Writing the nuclear potentials in terms of a multipole expansion, lets us differentiate the nuclear electromagnetic moments from structure deformation, and also gives us a clue about the ground state nuclear spin that are apt for the measurement of the nuclear electromagnetic moments.
A multipole expansion can be written for both the scalar and vector potentials \cite{Jackson1998-ga}, where the scalar potential, $\phi$, is linked to the charge distribution, $\rho$, by $\phi(\vec{R}) = 1/(4\pi\epsilon_0)\int d^3 \vec{r}\cdot\{\rho(\vec{r})/|\vec{R}-\vec{r}|\}$, and the vector potential, $\vec{A}$, is linked to the current density, $\vec{j}$, by $\vec{A}(\vec{R}) = \mu_0/(4\pi)\int d^3 \vec{r}\cdot\{\vec{j}(\vec{r})/|\vec{R}-\vec{r}|\}$~($\epsilon_0=8.8541878128(13)\times10^{-12}~$F/m and $\mu_0=1.25663706212(19)\times10^{-6}~$N/A$^2$ are the vacuum permittivity and permeability, respectively \cite{Tiesinga2021-qm}). Here, the scalar potential satisfies $\vec{E}=-\vec{\nabla}\phi(\vec{R})$, while the vector potential satisfies $\vec{B}=\vec{\nabla}\times\vec{A}(\vec{R})$, where $\vec{E}$ and $\vec{B}$ are electric and magnetic fields, respectively.

Both, scalar and vector potentials can be written as a sum of $2^{l}$-pole moments. From the Purcell-Ramsey-Schiff theorem \cite{Purcell1950-xw,Schiff1963-bz} the nuclear scalar potential can be written as a Laplace expansion \cite{Griffiths2004-dj},
\begin{eqnarray}
\phi({\bf R})&=&\frac{1}{4\pi\epsilon_0}\sum^{\infty}_{l=0} \frac{1}{R^{l+1}}\sum^{l}_{m=-l} (-1)^m Y^m_l(\hat{\vec{R}}) \mathcal{Q}^m_l,\label{eq1}\\
\mathcal{Q}^m_l&=&\sqrt{\frac{4\pi}{2l+1}} \int d^3 \vec{r}\cdot\rho(\vec{r}) r^l Y^m_l\left(\frac{\vec{r}}{r}\right),\label{eq1b}
\end{eqnarray}
where, $\mathcal{Q}^m_l$ is the electric $2^{l}$-pole moment, and $Y^m_l$ are the spherical harmonics. In a rotationally symmetric system, when canonically considering the z-axis to be the direction along the spin, the only relevant multipole moment corresponds to $m=0$. The nuclear charge, electric dipole moment (EDM: $d$), and electric quadrupole moment (EQM: $Q$), respectively, are given by \cite{Spevak1997-bg}:
\begin{eqnarray}
\mathcal{Q}^{m=0}_{l=0} &=& \int d^3 \vec{r}\cdot\rho(\vec{r}) = Z \mathbb{e}~\label{eq2-0},\\
\mathcal{Q}^{m=0}_{l=1} &=& \int d^3 \vec{r}\cdot~r\cos(\theta)~\rho(\vec{r}) = d~\label{eq2-1},\\
\mathcal{Q}^{m=0}_{l=2} &=& \int d^3 \vec{r}\cdot~r^2 \left(3\cos^2(\theta)-1\right)\frac{1}{2}~\rho(\vec{r})=Q,~\label{eq2-2}
\end{eqnarray}
where $Z$ is the atomic number of the nucleus, and $\mathbb{e}$ is the elementary charge. Similarly, the vector potential can be expanded as \cite{Gray1976-mr}:
\begin{eqnarray}
\vec{A}(\vec{R})&=&\frac{\mu_0}{4\pi}\sum^{\infty}_{l=0}\frac{1}{R^{l+1}}\sum^{l}_{m=-l} (-1)^m Y^m_l(\hat{\vec{R}}) \mathcal{M}^m_l, \label{eq3}\\
\mathcal{M}^m_l&=&\frac{1}{l+1}\sqrt{\frac{4\pi}{2l+1}} \int d^3 \vec{r}\cdot  r^l Y^m_l\left(\frac{\vec{r}}{r}\right)\cdot\left[\vec{\nabla}\times\vec{j}(\vec{r})\right], \label{eq3b}
\end{eqnarray}
where $\mathcal{M}^m_l$ is the magnetic $2^{l}$-pole moment. Along the lines of Eqs.~\ref{eq2-0}-\ref{eq2-2}, the magnetic monopole moment, magnetic dipole moment (MDM: $\mu$), and magnetic quadrupole moment (MQM: $M$) terms can be written as \cite{Gray2010-fw}:
\begin{eqnarray}
\mathcal{M}^{m=0}_{l=0} &=& 0~\label{eq3-0},\\
\mathcal{M}^{m=0}_{l=1} &=& \int d^3 \vec{r}\cdot~\frac{1}{2}\vec{r}\times\vec{j}(\vec{r}) = \mu~\label{eq3-1},\\
\mathcal{M}^{m=0}_{l=2} &=& \int d^3 \vec{r}\cdot~\frac{1}{6}\left[\left(\vec{r}\times\vec{j}(\vec{r})\right)\vec{r}+\vec{r}\left(\vec{r}\times\vec{j}(\vec{r})\right)\right] = M.~\label{eq3-2}
\end{eqnarray}
A typical source for the origin of the nuclear current density, $\vec{j}(\vec{r})$, is its spin, $s$.

It is important to note that the spherical harmonics are mutually orthogonal. Since the expectation value is proportional to the square of the wave function, cases where the wave function has a spherical harmonic term, $Y^m_{l=s}$, forces all terms involving $Y^m_{l>2s}$ in Eqs.~\ref{eq1} and \ref{eq3} to zero. Accordingly, following the Wigner-Eckart theorem \cite{Hecht2000-fa}, from Eqs.~\ref{eq1}-\ref{eq1b} and \ref{eq3}-\ref{eq3b} it can then be seen that nuclei with spin $s$, can have up to and including electromagnetic $2^{2s}-$pole moments. Therefore we need at least a spin $1/2$ nucleus to posses a non-zero EDM and MDM, and at least a spin $1$ nucleus to posses a non-zero EQM and MQM.

Furthermore, it an be seen in Table~\ref{tab1} and its associated text, that in a nucleus with appropriately large MDM and non-zero spin, while EDM is CP violating, MDM obeys CP. Generalizing this along the lines of all electromagnetic $2^{l}$-pole moments: electric $2^{2\mathcal{N}}$-pole moments and magnetic $2^{2\mathcal{N}+1}$-pole moments obey CP, and electric $2^{2\mathcal{N}+1}$-pole moments and magnetic $2^{2\mathcal{N}}$-pole moments violate CP, where $\mathcal{N}$ is a whole  number, $\mathcal{N}\in\mathbb{N}$. Therefore, electromagnetic moments like EDM (electric dipole ($2^{2(0)+1}$-moment) and MQM (magnetic quadrupole ($2^{2(1)}$-moment), in the context of quantifying CP violation, are extremely attractive \cite{Bernreuther1991-xl}. In this work, we are concerned with nuclei that have a ground state spin of at least $1$, so that we can study their CP violating permanent nuclear MQM, and its relative contribution to atomic EDM.

\section{Nuclear Structure Deformation}

The nucleus can be non-spherical. The deviation from the perfectly spherical geometry is successfully modeled using a rotationally symmetric two-fluid liquid drop model \cite{Bohr1957-aw}, each fluid corresponding to the ensemble of protons or neutrons \cite{Bohr1953-we}. The even-even core of the nucleus contributes significantly to the structure of the nucleus \cite{Spevak1997-bg,Auerbach1996-sp}, which also allows us to treat most other cases effectively as single particle systems. Following from the hydrodynamic origins of the liquid drop model \cite{Weizsacker1935-pm,Bohr1939-lr,Gamow1997-cg}, deformation in nuclei is also described in terms of $2^{l}$-pole moments, similar to the electromagnetic moments in Eqs.~\ref{eq1b} and \ref{eq3b}. The surface of a rotationally symmetric deformed nucleus can be described by \cite{Myers1969-qj,Leander1986-fr,Nazarewicz1990-yr,Butler1991-os}
\begin{eqnarray}
R=c_V R_0 \left(1+\sum_{l=2}\beta_l Y^{m=0}_{l}\right),~\label{eq4}
\end{eqnarray}
where $c_V=1-\{(1/\sqrt{4\pi})\sum_{l=2}\beta^2_l\}$ is the volume normalization which ensures that the volume of the nucleus scales linearly with $A$ regardless of the deformation, $R_0=1.2~\text{fm}\cdot A^{1/3}$, $A$ is the total number of nucleons in the nucleus, and $\beta_l$ are the $2^l$-pole structure deformation coefficients.

In Eq.~\ref{eq4}, the imposition of rotational symmetry forces $m=0$, similar to the reason we only considered electromagnetic moments in Eqs.~\ref{eq2-0}-\ref{eq2-2} and \ref{eq3-0}-\ref{eq3-2} with $m=0$. It must be noted that the connection between nuclear spin and the electromagnetic moments arising from the Wigner-Eckart theorem is not applicable to nuclear structure deformation, \emph{ie.} even a spin-$0$ nucleus may and routinely does have deformations corresponding to $l\ge2$ (making their $\beta_{l\ge2}\ne0$).

\begin{figure}[t]
\centering
\begin{tabular}{cc}
    \multirow{-13.5}{*}{\includegraphics[width=0.44\textwidth]{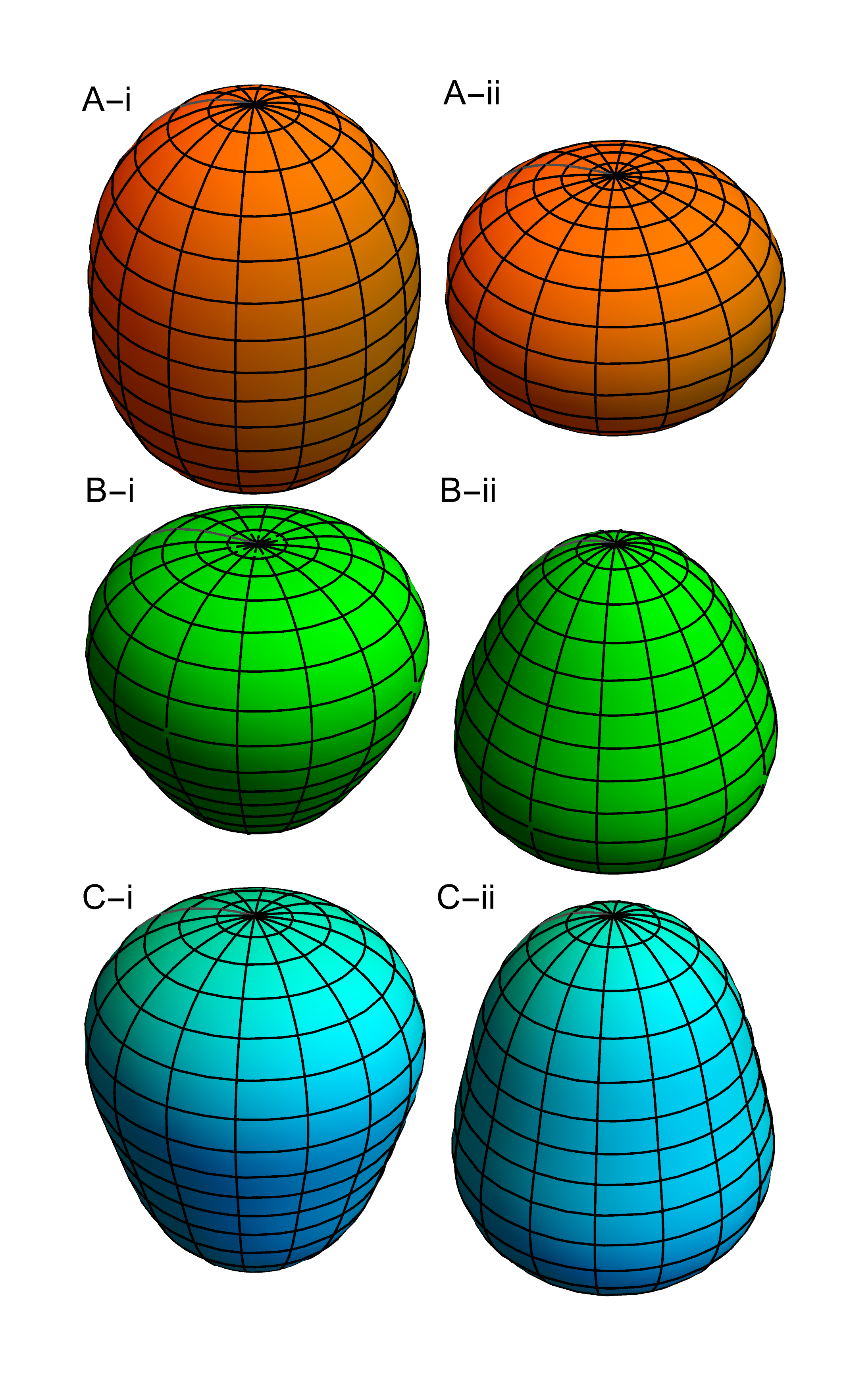}} & \includegraphics[width=0.54\textwidth]{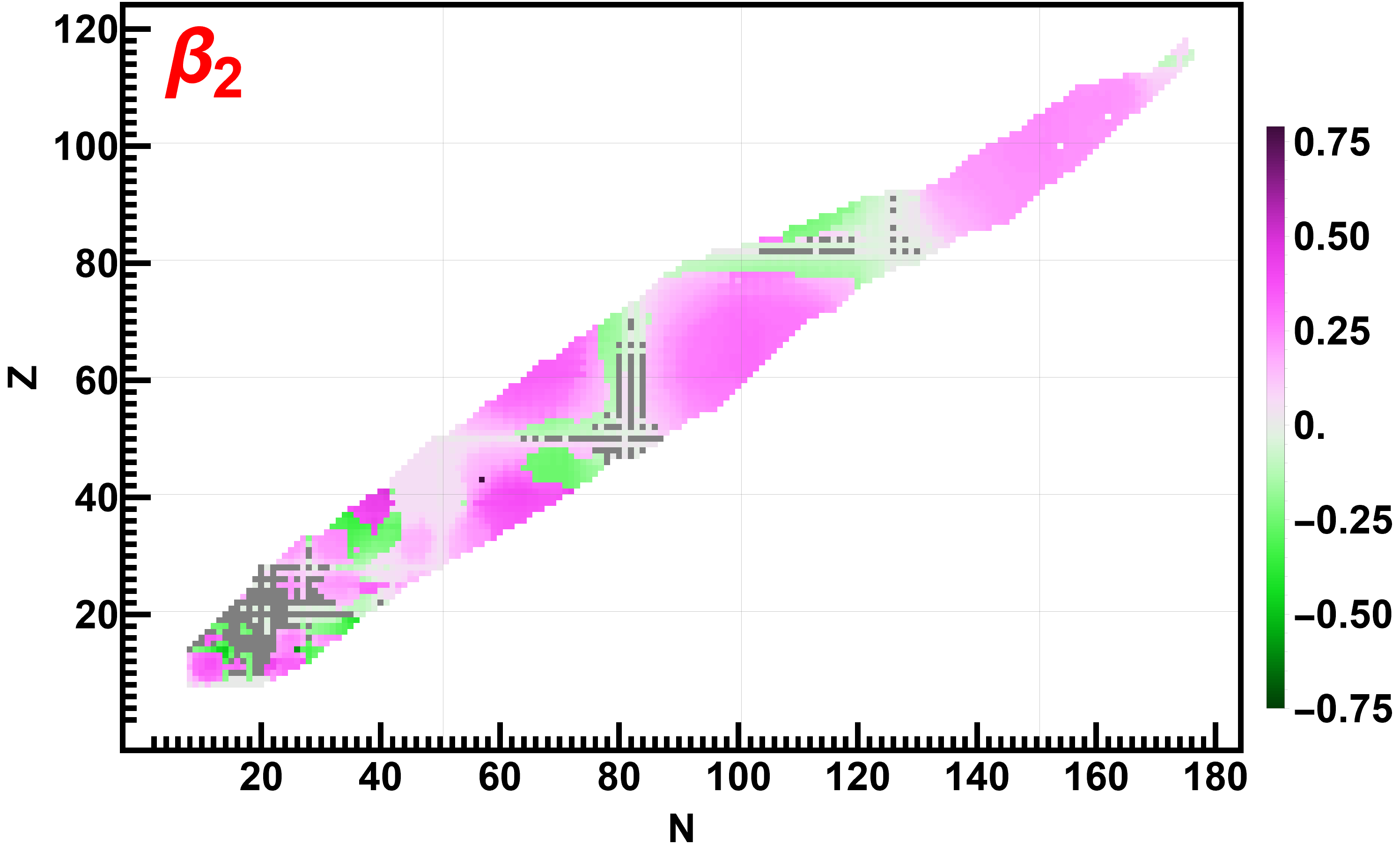} \\
  &  \includegraphics[width=0.54\textwidth]{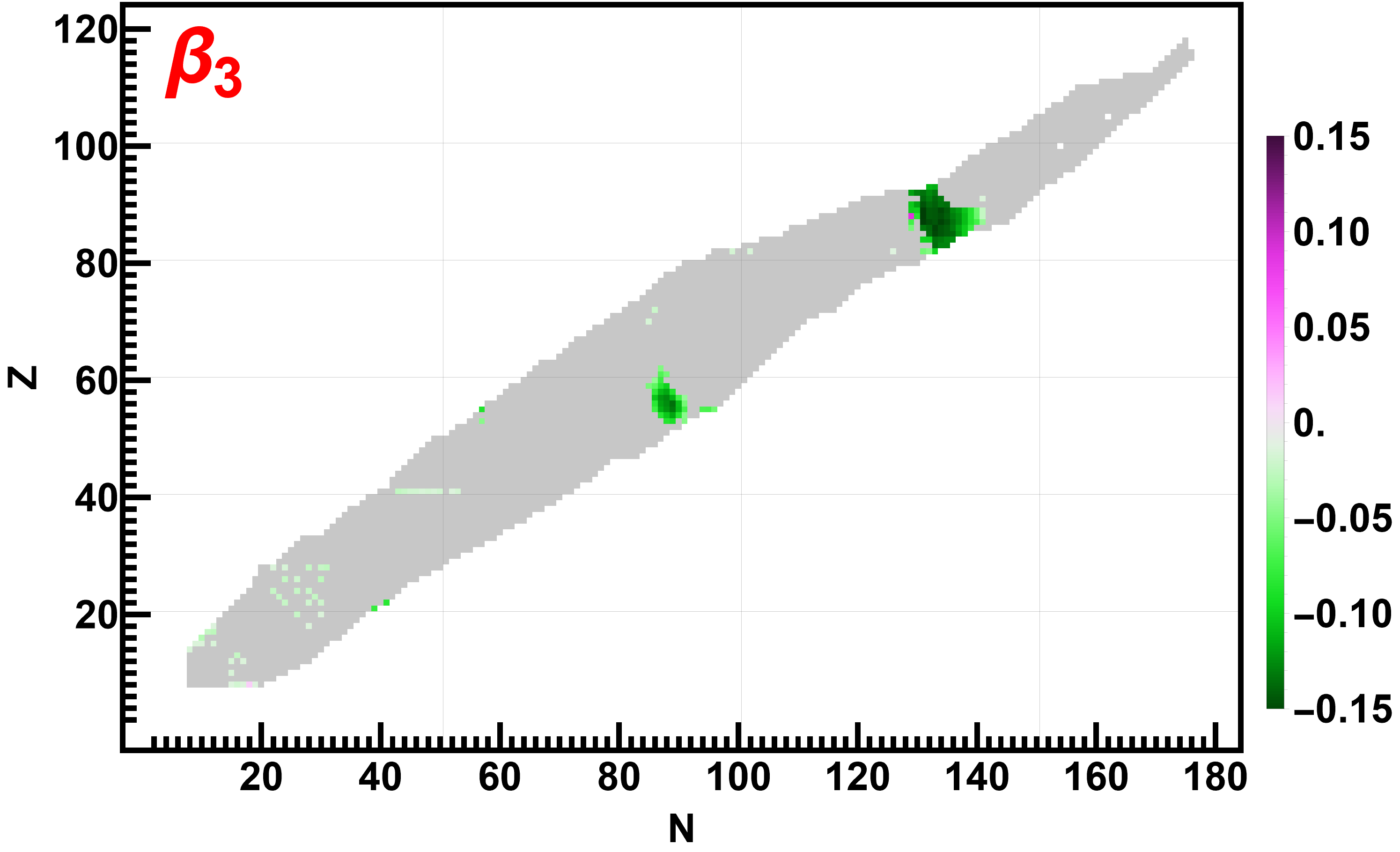}
\end{tabular}
\caption[]{[Left] Rendering of deformed nuclei - A: Quadrupole deformation (i: prolate, $\beta_2>0$; ii: oblate: $\beta_2<0$). B: Octupole deformation (i: $\beta_3>0$; ii: $\beta_3<0$). C: Prolate plus Octupole deformation (i: $\beta_2>0$, $\beta_3>0$; ii: $\beta_2>0$, $\beta_3<0$). [Right] Quadrupole (Top) and octupole (Bottom) deformation coefficients from the M\"oller-Nix model \cite{Moller1995-mx} for all isotopes whose binding energies are positive. Isotopes marked with light gray indicates that their binding energies are positive but their corresponding deformation is negligible.
}
\label{fig1}
\end{figure}

In this work, we have used structure deformation coefficients from the M\"oller-Nix model \cite{Moller1995-mx}. The M\"oller-Nix model is also based on the liquid drop model, but combines it with a folded Yukawa potential \cite{Moller1988-hp} in order to add shell model corrections using the Strutinsky method \cite{Strutinsky1967-hg,Strutinsky1968-ys}. With this combination, the M\"oller-Nix model is able to include the nucleon pairing effect \cite{Weizsacker1935-pm}, and most importantly, handle rotational degrees of freedom. The M\"oller-Nix model comprehensively provides deformation coefficients for all cases of even-even, even-odd, and odd-odd nuclei.

Examples of purely quadrupole and octupole deformed nuclei using Eq.~\ref{eq4} can be seen in Fig.~\ref{fig1} (Left) A and B, respectively. Here, for each of the two cases of quadrupole and octupole deformation, we have presented the shape of the nuclei for both positive (A-i: $\beta_2=+0.3$, B-i: $\beta_3=+0.2$) and negative (A-ii: $\beta_2=-0.3$, B-ii: $\beta_3=-0.2$) deformation coefficients. In the case of purely quadrupole deformed nuclei, Figs.~\ref{fig1} (Left) A-i and A-ii clearly demonstrate prolate and oblate shapes determined by the sign of the $\beta_2$ coefficient. In the case of purely octupole deformed nuclei, Figs.~\ref{fig1} (Left) B-i and B-ii clearly demonstrate the pear shapes determined by the $\beta_3$ coefficient. Since nuclei can be both quadrupole and octupole deformed simultaneously, we have also further shown a combination of prolate quadrupole deformation with both positive and negative octupole deformations in Figs.~\ref{fig1} (Left) C-i and C-ii, respectively.

Furthermore, we have plotted the $\beta_2$ and $\beta_3$ coefficient values from the M\"oller-Nix model in Figs.~\ref{fig1} (Right-Top) and (Right-Bottom), for the ground state of the isotopes, respectively. The collective behavior of both quadrupole and octupole deformation is clearly visible. Islands of quadrupole deformation are identifiable on either side of a magic number of nucleons, and the two major octupole deformation clusters, centered around $Z=\{56,87\}$ and $N=\{89,134\}$, are also visible. Canonically, the axis of symmetry in Figs.~\ref{fig1} (Left) A-C is also the axis of rotation. In case of non-zero spin, the spin vector sets the direction, where the positive direction is oriented along the same axis of symmetry. 
One of the poles, along the axis of symmetry, where the grid lines meet, is visible in each of these example deformed nuclei in Figs.~\ref{fig1} (Left). %
The predominance of prolate [over oblate] quadrupole deformation arises from quantum mechanics \cite{Zickendraht1990-cx}. Combining this insight with the direction set by the spin, and considering that negative parity states (with opposite $\beta_3$) in octupole deformed nuclei arise from vibration of quadrupole deformed states, makes evident the fact that most octupole deformed nuclei are characterized by a negative $\beta_3$ \cite{Butler2020-mx}. Therefore, Fig.~\ref{fig1} (Left) C-ii represents the most common ground state for each isotope with significant deformation seen in both Figs.~\ref{fig1} (Right-Top) and (Right-Bottom).

There are newer models which also calculate the deformation coefficients, \emph{eg. ref.} \cite{Agbemava2016-rx}, which uses covariant density functional theory, and \emph{ref.} \cite{Ebata2017-nt}, which combines a Skyrme Hartree-Fock model with Bardeen–Cooper–Schrieffer(BCS) type pairing. However, \emph{ref.} \cite{Agbemava2016-rx} only handles even-even nuclei, whereas \emph{ref.} \cite{Ebata2017-nt} lacks comprehensive calculation of octupole deformation for relevant species. When available, these models predict deformation coefficients that are mutually consistent, within their model uncertainties \cite{Mohanmurthy2020-np}.

\section{Atomic Electric Dipole Moment due to Nuclear Magnetic Quadrupole Moment}

Typically, atomic EDMs are much smaller than nucleon EDMs, owing to the screening of the nuclear EDM by the electron cloud \cite{Schiff1963-bz}. However, the electron screening is imperfect when \cite{Liu2007-wk}: (i) the valence electrons are highly relativistic, \emph{eg.} in paramagnetic atoms like $^{210}$Fr \cite{Engel2013-rk}; (ii) the nucleus is heavily quadrupole and octupole deformed, \emph{eg.} in diamagnetic atoms like $^{225}$Ra \cite{Dzuba2002-or}; or (iii) there are significant CP violating interactions between the electrons and the nucleus, \emph{eg.} in diamagnetic atoms like $^{199}$Hg \cite{Chupp2015-ns}. The residual CP violating electromagnetic moment of a nucleus is usually referred to as the nuclear Schiff moment.

\sloppy Furthermore, the nuclear MQM is not screened by the electron cloud, thereby dramatically increasing its relative contribution to the atomic EDM \cite{Flambaum1997-ae}. While the nuclear Schiff moment is enhanced by the simultaneous octupole and quadrupole deformations of the nucleus \cite{Spevak1997-bg,Auerbach1996-sp,Dzuba2002-or,Flambaum2019-oo}, the nuclear MQM is enhanced by quadrupole deformation of the nucleus alone \cite{Flambaum1994-fp,Lackenby2018-up}.

Nuclear MQMs along with nuclear Schiff moments, being CP violating, contribute to atomic EDMs \cite{Flambaum2014-qn,Lackenby2018-ua}. In this section, we are only concerned with the approximate trends that dictate the enhancement of nuclear MQMs based on their quadrupole deformation, and their contribution to atomic EDMs, without explicitly calculating them, analogous to nuclear Schiff moments and their contributions to atomic EDMs in \emph{ref.} \cite{Mohanmurthy2020-np}. The exact recipe for calculating MQMs of deformed nuclei is presented in \emph{refs.} \cite{Flambaum1994-fp,Lackenby2018-up}. 

The trends dictating the variation of nuclear MQMs as a function of nuclear quadrupole deformation (\emph{ref.}~\cite{Khriplovich2012-no}, Section 10.3), and atomic EDM due to the nuclear MQM \cite{Sushkov1984-cb} are:
\begin{eqnarray}
M\propto \beta_2 Z A^{2/3}~~;~~d_{\text{atom}}\propto \frac{M Z^2}{E_{+}-E_{-}}~~\implies~~d^{M}_{\text{atom}}\propto\frac{\beta_2 Z^3 A^{2/3}}{E_{+}-E_{-}},\label{eq5}
\end{eqnarray}
where $E_{\pm}$ are the energies of the ground state parity doublet. Even though \emph{ref.}~\cite{Khriplovich2012-no} (Section 10.3), indicates the trend $M\propto \beta_2 Z$, without the dependence on $A^{2/3}$, we have included this factor because the average rotational energy of a deformed nucleus follows the trend $\hbar \bar{\omega}\propto (45A^{-1/3}-25A^{-2/3})$ \cite{Lackenby2018-up,Brown2017-kh}, which itself relies on the trend that $\beta_2\propto A^{1/3}$ \cite{Khriplovich2012-no} (P. 197). We have verified that the quadrupole deformation coefficient, $\beta_2$, indeed follows this trend in the appendix Section~\ref{sec:A}. Therefore, the scaling of atomic EDM, $d^{M}_{\text{atom}}$, as a function of $Z^3A^{2/3}/(E_{+}-E_{-})$, is consistent with the same extracted via the nuclear Schiff moment \cite{Mohanmurthy2020-np}. The trends in Eq.~\ref{eq5} represent a relationship between electromagnetic moments, particularly the magnetic moments in Eqs.~\ref{eq3-0}-\ref{eq3-2}, to structure deformation coefficients in Eq.~\ref{eq4}. These trends are sufficient to survey the nuclide chart, in order to study the quadrupole deformed systems in which atomic EDM is maximally enhanced.

\section{Survey of Nuclear Quadrupole Deformation}

\begin{figure}[t]
\centering
\begin{tabular}{cc}
    \includegraphics[width=0.49\textwidth]{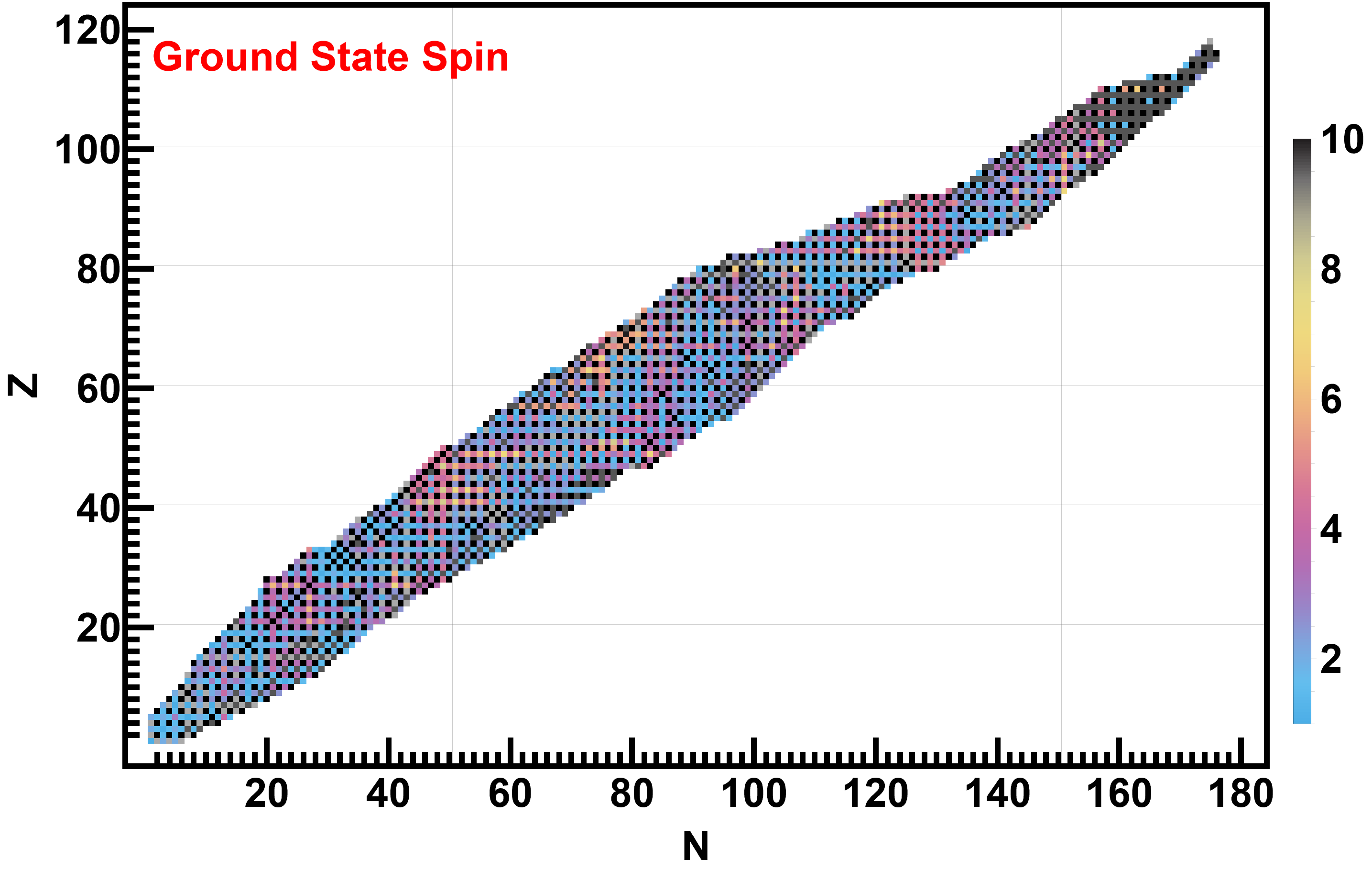} &
    \includegraphics[width=0.49\textwidth]{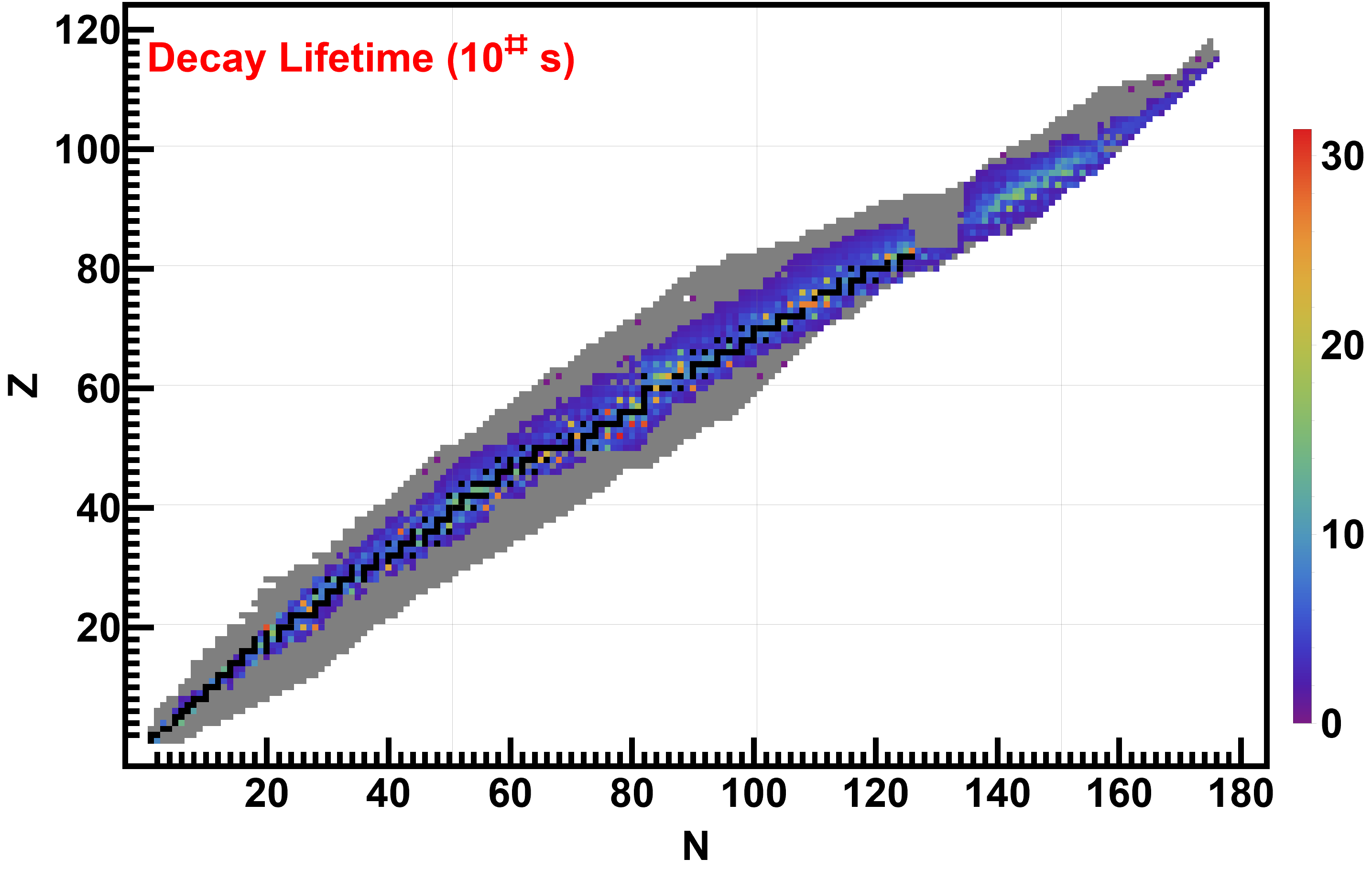} \\
    \multicolumn{2}{c}{\includegraphics[width=\textwidth]{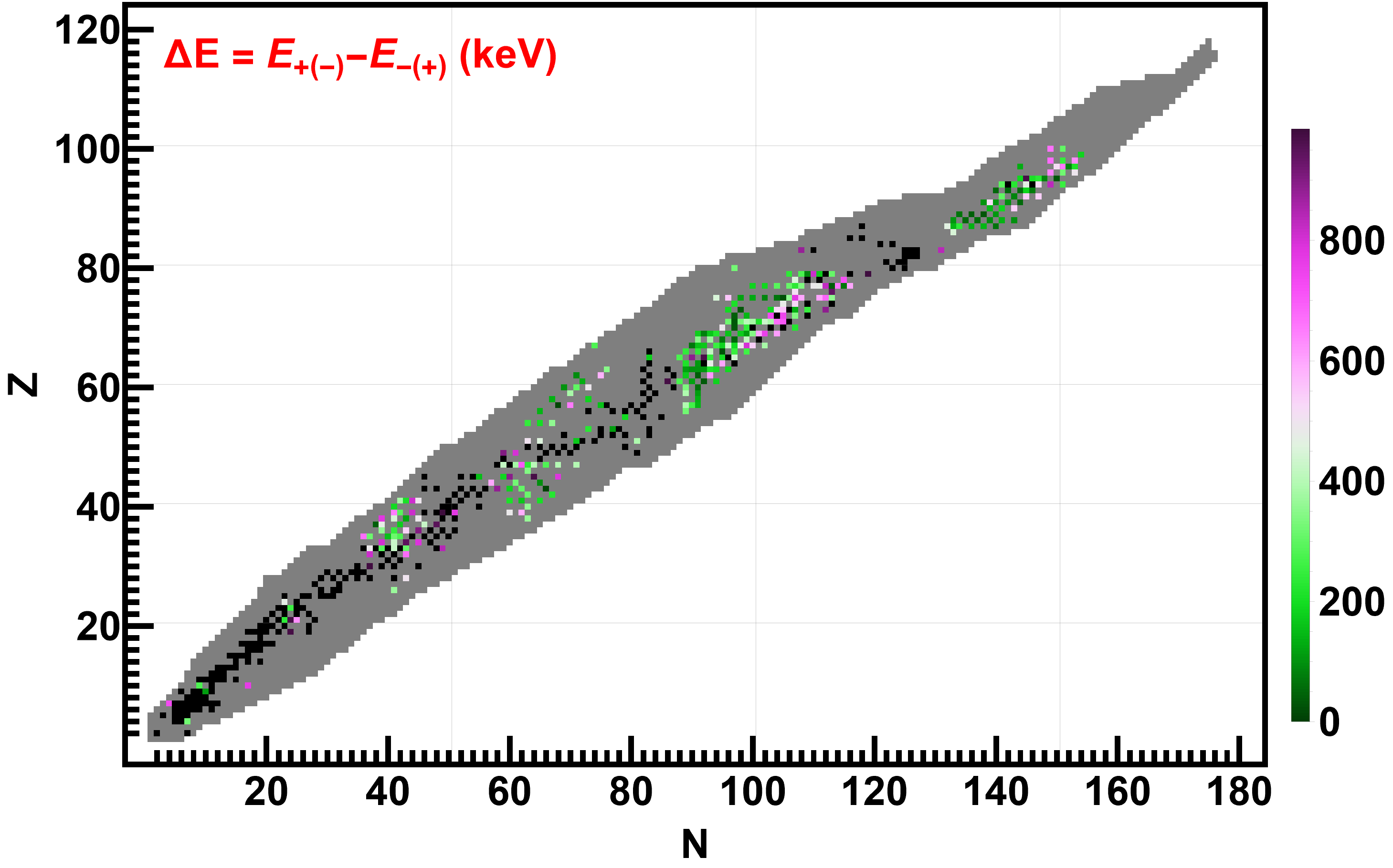}}
\end{tabular}
\caption[]{[Top-Left] Ground state spin of nuclei with positive binding energy. Black and gray colors represent spin: $|s|=0$ and spin: $|s|=1/2$ species, respectively. [Top-Right] Ground state decay lifetime of nuclei with positive binding energy, where the lifetime is $\ge1~$minute, represented by colors corresponding to seconds in power of $10$. [Bottom] Energy difference between the states that make up the ground state parity doublet in keV. Black represents that this energy difference is greater than $1~$MeV, and gray represents that this energy difference has not been measured.}
\label{fig3}
\end{figure}

As a part of this survey, we considered: (i) quadrupole deformation, shown in Fig.~\ref{fig1} (Right-Top), (ii) ground state spins, shown in Fig.~\ref{fig3} (Top-Left), (iii) ground state parity doublet energy difference, shown in Fig.~\ref{fig3} (Bottom), (iv) decay lifetime, shown in Fig.~\ref{fig3} (Top-Right), and (v) stopped beam rates at FRIB, of $3176$ isotopes whose binding energies are positive. For quadrupole deformation, we relied on the $\beta_2$ values in \emph{ref.} \cite{Moller1995-mx}. For the binding energies, nuclear ground state spins, and decay lifetimes, we relied on the National Nuclear Data Center (NNDC) database \cite{nndc}, and for the isomers of isotopes and their associated energy levels in order to determine the ground state parity doublet energy difference, we used the International Atomic Energy Agency - Nuclear Data Services (IAEA-NDS) \cite{iaea}, both of which are based on ENSDF databases \cite{ensdf}. For the stopped beam rates at FRIB, we used the ultimate yields from LISE$^{++}$ \cite{lisepp}.

In this survey, we are only concerned with the quadrupole deformation and the enhancement of the nuclear MQM due to it. Therefore, we only considered isotopes with a ground state nuclear spin greater than or equal to one, $j\ge1$. We also only considered isotopes whose lifetimes are at least of the order $1~$minute, making it comparable to \emph{ref.} \cite{Mohanmurthy2020-np}. Furthermore, according to Eq.~\ref{eq5}, the nuclear MQM and the atomic EDM due to the nuclear MQM are proportional to the inverse of the ground state parity doublet energy difference, and therefore we only considered isotopes where this is lower than $1~$MeV. Typically, the best candidates have a ground state parity doublet energy difference on the order of $10~$keV \cite{Mohanmurthy2020-np}.

The relative atomic EDMs follow the trend in Eq.~\ref{eq5}: $d^{M}\propto\beta_2 Z^3 A^{2/3}/\Delta E_{\pm}$. We present this trend in Table~\ref{tab2}, after normalizing them with that of $^{223}$Ra, in order to keep it consistent with \emph{ref.} \cite{Mohanmurthy2020-np}. Most experiments that measure EDMs in reality measure the precession frequency of the species by applying magnetic and electric fields, and employing the Ramsey technique of separated oscillating fields \cite{Purcell1950-xw,Ramsey1949-iz,Ramsey1950-ma}. The measurement uncertainty associated with the EDM using the Ramsey technique scales as the square root of the number of particles, $\sqrt{W^{\text{(FRIB)}}}$, where $W^{\text{(FRIB)}}$ is the ultimate stopped beam rate achievable at FRIB. Ultimately, we are concerned with the combined impact of a highly enhanced MQM contribution according to Eq.~\ref{eq5}, and access to a high experimental sensitivity. So we have also indicated an impact factor, defined by $\mathcal{I}\propto\{\beta_2 Z^3 A^{2/3}/\Delta E_{\pm}\}\cdot \sqrt{W^{\text{(FRIB)}}}$, in Table~\ref{tab2}, after normalizing them against $^{223}$Ra. Since atomic EDM due to nuclear MQM scales as $\propto Z^3$, heavy elements in this list are typically more impactful. 

The isotope with the best relative atomic EDM, arising from its nuclear MQM enhanced by its quadrupole deformation, is $^{245}$Am. 
In Table~\ref{tab2}, we have shown all candidate isotopes, which have a relative atomic EDM that are up to a factor of $10$ lower than $^{245}$Am. However, the beam rates for trans-Neptunium nuclei are not available in \emph{ref.}~\cite{lisepp}. We used the same LISE$^{++}$ toolkit \cite{Tarasov2002-jy,Bazin2002-gk,Tarasov2004-yc,Tarasov2016-wr,Tarasov2008-cr,Kuchera2016-zw} as in \emph{ref.}~\cite{lisepp}, using a $5~$kW - $177~$MeV/u beam of $^{238}$U incident on the FRIB production target \cite{Avilov2015-ja,Pellemoine2014-zl}. This new calculation of the beam rates, for the isotopes of $^{239,241,242,243,245}$Am and $^{237}$Pu, have also been presented in Table~\ref{tab2} and are highlighted in red. Even though isotopes of Americium, like other trans-Uranium isotopes, suffer from small rates at FRIB, an EDM experiment could potentially be conceived off the beam-line. We have presented certain isotopes that have a relative atomic EDM that is within a factor of $10$ of $^{245}$Am, but whose rates are below $0.1$/day, in Table~\ref{tab3}.

In Table~\ref{tab2}, some isotopes have a lifetime of a few minutes, while others are on the order of many thousand years. Statistical precision of a measured EDM using the Ramsey technique of separated oscillating fields scales as the inverse of measurement time. The measurement time is usually restricted by the typical spin-coherence time in experiments on the order of a few minutes. Our choice of constraint, set to $1~$minute on the decay lifetime, is long enough so that the factor coming from decay lifetimes can be comfortably left out.

\begin{table}[p]
\renewcommand{\arraystretch}{1.2}%
\setlength{\tabcolsep}{2.5pt}
\centering
\caption{Deformed nuclei whose production rates at FRIB are finite, with their relative theoretical expectation of EDM ($d^{M}$, Eq.~\ref{eq5}), ultimate beam rate at FRIB ($W^{(\text{FRIB})}$), and their relative impact ($\mathcal{I}\propto d^{M}\sqrt{W^{(\text{FRIB})}}$). Their relative theoretical expectation of EDM and their relative impact, have both been normalized to $^{223}$Ra. Their quadrupole deformation coefficients, $\beta_2$, are from the M\"oller-Nix model \cite{Moller1995-mx}. Units of beam rate, $W^{(\text{FRIB})}$ is $10^6/$s \cite{Tarasov2002-jy,Bazin2002-gk,Tarasov2004-yc,Tarasov2016-wr,Tarasov2008-cr,Kuchera2016-zw}. Parity doublet energy splitting, $\Delta E = E_--E_+$, is in units of keV. $^{*}$: Parity doublet energy levels have not been adopted by the National Nuclear Data Center (NNDC) database \cite{nndc}. 
} \label{tab2}
\begin{tabular}{llrrcrcll}
\hline
&& \multicolumn{1}{c}{$T_{1/2}$} & \multicolumn{1}{c}{$\beta_2$} & \multicolumn{1}{c}{$j^{\pi}$} & \multicolumn{1}{c}{$\Delta E$(keV)} & $|d^{M}|$ & $W^{(\text{FRIB})}$ & $\sim|\mathcal{I}|$\\
\hhline{=========}
$Z=60$ & $^{151}$Nd & $12.44(7)$ min & $0.252$ & $3/2^+$ & $57.6741(4)$ & $0.34$ & $4.06$ & $0.12$\\
\hline
$Z=61$ & $^{153}$Pm & $5.25(2)$ min & $0.270$ & $5/2^-$ & $32.194(10)$ & $0.70$ & $3.58$ & $0.24$\\
\hline
$Z=62$ & $^{153}$Sm & $46.284(4)$ hour & $0.261$ & $3/2^+$ & $35.844(3)$ & $0.64$ & $17.1$ & $0.48$\\
\hline
$Z=67$ & $^{157}$Ho & $12.6(2)$ min & $0.235$ & $7/2^-$ & $66.911(20)$ & $0.39$ & $228$ & $1.07$\\
\hline
\rowcolor{yellow!50}$Z=68$ & $^{163}$Er & $75.0(4)$ min & $0.272$ & $5/2^-$ & $69.23(1)$ & $0.47$ & $56.3$ & $0.64$\\
\rowcolor{yellow!50}& $^{165}$Er & $10.36(4)$ hour & $0.282$ & $5/2^-$ & $47.159(4)$ & $0.73$ & $18.4$ & $0.57$\\
\hline
\rowcolor{yellow!50}$Z=69$ & $^{161}$Tm & $30.2(8)$ min & $0.254$ & $7/2^+$ & $78.20(3)$ & $0.40$ & $400$ & $1.45$\\
\rowcolor{yellow!50}& $^{168}$Tm & $93.1(2)$ day & $0.294$ & $3^+$ & $112(1)$ & $0.34$ & $21.3$ & $0.28$\\
\hline
\rowcolor{yellow!50}$Z=70$ & $^{167}$Yb & $17.5(2)$ min & $0.274$ & $5/2^-$ & $29.656(8)$ & $1.23$ & $164$ & $2.86$\\
\hline
$Z=72$ & $^{169}$Hf & $3.24(4)$ min & $0.274$ & $5/2^-$ & $38.18(4)$ & $1.05$ & $599$ & $4.66$\\
\hline
$Z=73$ & $^{171}$Ta & $23.3(3)$ min & $0.255$ & $(5/2^+)$ & $31.2(1)$ & $1.26$ & $710$ & $6.09$\\
& $^{180}$Ta & $8.154(6)$ hour & $0.269$ & $1^+$ & $130.38(7)$ & $0.33$ & $88.3$ & $0.56$\\
\hline
$Z=75$ & $^{173}$Re & $1.98(26)$ min & $0.218$ & $(5/2^-)$ & $118.1(12)$ & $0.31$ & $257$ & $0.90$\\
& $^{175}$Re & $5.89(5)$ min & $0.238$ & $(5/2^-)$ & $125.3(1)$ & $0.32$ & $559$ & $1.37$\\
& $^{177}$Re & $14(1)$ min & $0.238$ & $5/2^-$ & $84.70(10)$ & $0.48$ & $831$ & $2.51$\\
& $^{179}$Re & $19.5(1)$ min & $0.239$ & $5/2^+$ & $65.35(9)$ & $0.63$ & $905$ & $3.44$\\
& $^{180}$Re & $2.46(3)$ min & $0.248$ & $1^-$ & $20.10(10)$ & $2.13$ & $852$ & $11.28$\\
\hline
$Z=77$ & $^{190}$Ir & $11.78(10)$ day & $0.164$ & $4^-$ & $36.154(25)$ & $0.88$ & $1480$ & $6.14$\\
& $^{192}$Ir & $73.829(11)$ day & $0.145$ & $4^+$ & $66.830(20)$ & $0.42$ & $1210$ & $2.65$\\
\hline
$Z=79$ & $^{188}$Au & $8.84(6)$ min & $-0.156$ & $1^{(-)}$ & $82.70(8)$ & $0.39$ & $78.2$ & $0.63$\\
\hline
\rowcolor{yellow!50}$Z=87$ & $^{223}$Fr & $22.00(7)$ min & $0.146$ & $3/2^{(-)}$ & $ 134.48(4)$ & $0.34$ & $3.60$ & $0.12$\\
\rowcolor{yellow!50}& $^{225}$Fr & $3.95(14)$ min & $0.163$ & $3/2^-$ & $142.59(3)$ & $0.36$ & $0.92$ & $0.06$\\
\hline
\rowcolor{yellow!50}$Z=88$ & $^{223}$Ra & $11.43(5)$ day & $0.156$ & $3/2^+$ & $50.128(9)$ & $1.00$ & $30.4$ & $1.00$\\
\rowcolor{yellow!50}& $^{227}$Ra & $42.2(5)$ min & $0.181$ & $3/2^+$ & $90.0343(17)$ & $0.65$ & $3.40$ & $0.22$\\
\rowcolor{yellow!50}& $^{229}$Ra & $4.0(2)$ min & $0.189$ & $5/2^+$ & $ 137.45(6)$ & $0.45$ & $0.68$ & $0.07$\\
\rowcolor{yellow!50}& $^{231}$Ra & $103.9(14)$ s & $0.207$ & $(5/2^+)$ & $ 95.50(9)$ & $0.71$ & $69.2m$ & $0.03$\\
\hline
$Z=89$ & $^{223}$Ac & $2.10(5)$ min & $0.147$ & $(5/2^-)$ & $64.62(4)$ & $0.76$ & $152$ & $1.70$\\
& $^{225}$Ac & $9.920(3)$ day & $0.164$ & $(3/2^-)$ & $40.10(4)$ & $1.37$ & $84.4$ & $2.28$\\
& $^{227}$Ac & $21.772(3)$ year & $0.172$ & $3/2^-$ & $27.369(11)$ & $2.11$ & $38.7$ & $2.38$\\
& $^{229}$Ac & $62.7(5)$ min & $0.189$ & $(3/2^+)$ & $104.3(4)$ & $0.61$ & $13.6$ & $0.41$\\
\hline
$Z=90$ & $^{229}$Th & $7880(120)$ year & $0.190$ & $5/2^+$ & $146.3569(14)$ & $0.45$ & $136$ & $0.95$\\
& $^{231}$Th & $25.57(8)$ hour & $0.198$ & $5/2^+$ & $185.7160(13)$ & $0.38$ & $61.7$ & $0.54$\\
\hline
$Z=91$ & $^{229}$Pa$^*$ & $1.50(5)$ day & $0.190$ & $(5/2^+)$ & $0.06(5)$ & $1145$ & $20.7$ & $945$\\
& $^{231}$Pa & $3.276(11)\times10^4$ year & $0.198$ & $3/2^-$ & $102.2685(21)$ & $0.70$ & $548$ & $2.97$\\
& $^{233}$Pa & $26.975(13)$ day & $0.207$ & $3/2^-$ & $94.660(11)$ & $0.80$ & $359$ & $2.75$\\
& $^{235}$Pa & $24.4(2)$ min & $0.215$ & $(3/2^-)$ & $51.79(17)$ & $1.53$ & $184$ & $3.76$\\
\hline
$Z=92$ & $^{235}$U & $7.04(1)\times10^8$ year & $0.215$ & $7/2^-$ & $81.724(4)$ & $1.00$ & $1610$ & $7.28$\\
\hline
$Z=93$ & $^{233}$Np$^*$ & $36.2(1)$ min & $0.207$ & $(5/2^+)$ & $\approx 50$ & $1.62$ & $14.6$ & $1.12$\\
& $^{235}$Np & $396.1(12)$ day & $0.215$ & $5/2^+$ & $49.10(10)$ & $1.72$ & $16.9$ & $1.28$\\
& $^{237}$Np & $2.144(7)\times10^6$ year & $0.215$ & $5/2^+$ & $59.54092(10)$ & $1.43$ & $16.8$ & $1.06$\\
& $^{238}$Np & $2.099(2)$ day & $0.215$ & $2^+$ & $182.8775(18)$ & $0.47$ & $15.8$ & $0.34$\\
& $^{239}$Np & $2.356(3)$ day & $0.223$ & $5/2^+$ & $74.6640(10)$ & $1.19$ & $773m$ & $0.19$\\
\hline
\rowcolor{red!25}$Z=94$ & $^{237}$Pu & $45.64(4)$ day & $0.215$ & $7/2^-$ & $224.25(5)$ & $0.39$ & $31.0m$ & $0.01$\\
\hline
\rowcolor{red!25}$Z=95$ & $^{239}$Am & $11.9(1)$ hour & $0.215$ & $(5/2)^-$ & $187.1(5)$ & $0.49$ & $12.2\mu$ & $0.31m$\\
\rowcolor{red!25}& $^{241}$Am & $432.6(6)$ year & $0.223$ & $5/2^-$ & $205.883(10)$ & $0.46$ & $563n$ & $62.6\mu$\\
\rowcolor{red!25}& $^{242}$Am & $16.02(2)$ hour & $0.224$ & $1^-$ & $230.527(3)$ & $0.42$ & $1.21\mu$ & $83.8\mu$\\
\rowcolor{red!25}& $^{243}$Am & $7364(22)$ year & $0.224$ & $5/2^-$ & $84.00(16)$ & $1.14$ & $0.243\mu$ & $0.11m$\\
\rowcolor{red!25}& $^{245}$Am & $2.05(1)$ hour & $0.224$ & $5/2^+$ & $28.27(13)$ & $3.45$ & $3.88p$ & $1.23\mu$\\
\hhline{=========}
\end{tabular}
\end{table}

\begin{table}[t]
\renewcommand{\arraystretch}{1.2}%
\setlength{\tabcolsep}{2.5pt}
\centering
\caption{Deformed nuclei whose production rates at FRIB are negligible, with the relative theoretical expectation for the EDM ($d^{M}$, Eq.~\ref{eq5}). Their relative theoretical expectation for the EDM has been normalized to $^{223}$Ra. Their quadrupole deformation coefficients, $\beta_2$, are from the M\"oller-Nix model \cite{Moller1995-mx}. Parity doublet energy splitting, $\Delta E = E_--E_+$, is in units of keV.}
\label{tab3}
\begin{tabular}{llrrrrc}
\hline
&& \multicolumn{1}{c}{$T_{1/2}$} & \multicolumn{1}{c}{$\beta_2$} & \multicolumn{1}{c}{$j^{\pi}$} & \multicolumn{1}{c}{$\Delta E$(keV)} & $d^{M}$\\
\hhline{=======}
$Z=97$ & $^{241}$Bk & $4.6(4)$ min & $0.224$ & $(7/2^+)$ & $128(7)$ & $0.79$ \\
& $^{249}$Bk & $330(4)$ day & $0.235$ & $7/2^+$ & $82.599(13)$ & $1.32$ \\
& $^{250}$Bk & $3.212(5)$ hour & $0.235$ & $2^-$ & $211.82(1)$ & $0.52$ \\
\hline
$Z=98$ & $^{249}$Cf & $351(2)$ year & $0.235$ & $9/2^-$ & $243.13(7)$ & $0.46$ \\
\hline
$Z=99$ & $^{253}$Es & $20.47(3)$ day & $0.236$ & $7/2^+$ & $181.3(5)$ & $0.65$\\
\hhline{=======}
\end{tabular}
\end{table}

Cooling and trapping the atoms and molecules helps achieve a high number density, increasing the statistical sensitivity of the measurement. Laser based cooling is one the chief methods. It needs a (nearly) closed transition, which satisfies appropriate quantum selection rules like the $6S_{1/2}(F=4)-6P_{3/2}(F=5)$ transition in Cesium, that can also be accessed by a commercially available laser, or in combination with secondary optics, like frequency doublers. Laser cooling has been traditionally demonstrated in alkali (Li, Na, K, Rb, Cs, Fr) and alkali-Earth (Mg, Ca, Sr, Ba, Ra \cite{Bishof2016-wu}) atoms, as well as in Nobel gases (He, Ne, Ar, Kr, Xe) \cite{McClelland2016-od}. Similarly, transition metals (Hg, Ag, Cd, but also Cr), P-block elements (Al, Ga, In), F-block Lanthanides (Dy, Ho, Er, Tm, Yb), certain aspects of whose atomic structure are comparable to the traditional candidates, have also been laser cooled \cite{McClelland2016-od}. Atoms which have been laser cooled are particularly attractive to future work and have been highlighted in yellow in Table~\ref{tab2}.

\section{Conclusion}

The field of measuring EDMs in atoms is a thriving one with measurements already made in paramagnetic atoms of $^{89}$Rb \cite{Ensberg1967-kt}, $^{133}$Cs \cite{Murthy1989-ay}, $^{171}$Yb \cite{Zheng2022-bh} and $^{205}$Tl \cite{Commins1994-fe}, and diamagnetic atoms of $^{199}$Hg\cite{Graner2016-ge} and $^{129}$Xe \cite{Allmendinger2019-el}. There is only one measurement made in an octupole deformed system: $^{225}$Ra \cite{Bishof2016-wu}. There are however proposals to measure the EDMs of $^{210}$Fr \cite{Inoue2015-kc} and $^{223}$Rn \cite{Tardiff2014-th}. Similarly, measurements in molecules also exist: HfF$^+$ \cite{Roussy2023-vd}, ThO \cite{ACME_Collaboration2014-et}, YbF \cite{Hudson2011-wy}, PbO \cite{Eckel2013-gv}, TlF \cite{Cho1991-vt}, as well as proposals to make measurements in BaF \cite{Flambaum2014-qn}, TaN \cite{Skripnikov2015-jl}, ThF$^+$ \cite{Skripnikov2015-hx,Loh2013-jj}, PbF \cite{Skripnikov2015-kn,Petrov2013-cb,Skripnikov2014-bm,McRaven2008-nq}, WC \cite{Lee2009-aq,Lee2013-vl}, RaO \cite{Flambaum2008-up,Kudashov2013-se}, RaF \cite{Isaev2012-jm,Kudashov2014-ye}, PtH$^+$ \cite{Meyer2006-dl,Skripnikov2009-ub}, HgX \cite{Prasannaa2015-df}, and FrAg \cite{Klos2022-zw}. Furthermore, in addition to the atoms where laser cooling has been achieved, molecules including MgF, AlF, CaF, SrF, YbF, BaF, RaF, TlF, BH, CH, BaH, AlCl, CaOH, SrOH, YbOH, and CaOCH$_3$ have also been laser cooled, paving the way for similar techniques being exploited in molecular systems \cite{Fitch2021-yy}.

It is important to note that the impact factor, $\mathcal{I}$, in Table~\ref{tab2} is only relevant if one were to conceive of an in-beam EDM experiment at FRIB. However, the $^{225}$Ra-EDM experiment \cite{Bishof2016-wu} was an off-line measurement, using a stand alone source of $^{225}$Ra atoms. Likewise the planned molecular experiments involving FrAg and RaAg \cite{Klos2022-zw} also plan to conduct an off-line measurement using a stand alone atomic beam source. Other considerations like the efficiency of their cooling and the cross-section with which these isotopes may form molecules with accessible transitions may dictate the relevant number density in these off-line measurements.

We here surveyed over $3176$ isotopes, and particularly searched for candidates with the lowest energy difference between the states that make up their ground state parity doublet and highest quadrupole deformation. We thus arrived at $48$ isotopes: $^{151}$Nd, $^{153}$Pm, $^{153}$Sm, $^{157}$Ho, $^{163,165}$Er, $^{161,168}$Tm, $^{167}$Yb, $^{169}$Hf, $^{171,180}$Ta, $^{173,175,177,179,180}$Re, $^{190,192}$Ir, $^{188}$Au, $^{223,225}$Fr, $^{223,227,229,231}$Ra, $^{223,225,227,229}$Ac, $^{229,231}$Th, $^{229,231,233,235}$Pa, $^{235}$U, $^{233,235,237-239}$Np, $^{237}$Pu, and $^{239,241-243,245}$Am,
listed in Table~\ref{tab2} which are the most primed for a measurement of an atomic EDM, given their highly enhanced nuclear MQM and large production rates at FRIB. Similarly, we also identified the five trans-Uranium isotopes listed in Table~\ref{tab3} which are of interest: $^{241,249,250}$Bk, $^{249}$Cf, and $^{253}$Es, 
but would need to be produced at an alternate facility due to their negligible rates at FRIB. A similar study in \emph{ref.} \cite{Mohanmurthy2020-np} surveyed the octupole deformations and identified a number of species among which are $^{223,225}$Fr, $^{223}$Ra, $^{223,225,227}$Ac, $^{229}$Th, and $^{229}$Pa that overlap with the list presented here (Table~\ref{tab2}). From Figs.~\ref{fig1} (Right-Top) and (Right-Bottom), the prevalence of quadrupole deformation compared with octupole deformation yields more candidate isotopes in this study of quadrupole deformation compared to the study in \emph{ref.} \cite{Mohanmurthy2020-np}. Furthermore, laser cooling of $11$ isotopes $^{163,165}$Er, $^{161,168}$Tm, $^{167}$Yb, $^{223,225}$Fr, $^{223,227,229,231}$Ra
from the above list has already been demonstrated. Of all isotopes, the $3$ isotopes $^{223,225}$Fr and $^{223}$Ra provide tantalizingly interesting possibilities by the virtue of them having very close ground state parity doublets, maximal quadrupole and octupole deformation, as well as laser cooling already being demonstrated for them.

It was noted in \emph{ref.}~\cite{Mohanmurthy2020-np} that the deformation coefficients (in Figs.~\ref{fig1} (Right-Top) and (Right-Bottom)) sometimes differ by as much $\sim50\%$, depending on the model of choice. This variation was also implicitly observed in this work. In this survey of $3176$ isotopes, $90$ isotopes had ground state spins which did not match the theoretical value, assuming the deformation shown in Figs.~\ref{fig1} (Right-Top) and (Right-Bottom). In such cases, we used the experimentally measured values. The relativistic term in Eq.~\ref{eq5} has been neglected here, and thereby depreciates the relative enhancement of nuclear MQM in heavier isotopes. But this source of uncertainty is typically less than that due to the variation in nuclear structure coefficients. Lastly, the level scheme has not been well established in many isotopes. This impacts our survey as well, particularly in the cases of $^{229}$Pa and $^{233}$Np, as indicated in Table~\ref{tab2}. This work, especially in Eqs.~\ref{eq5}, relies on relative factors that contribute to the nuclear MQM and in turn the atomic EDM. However, in order to capture the entirety of the effects that contribute to the nuclear MQM, ultimately a full calculation of nuclear deformation using a modified Nilsson model and based on \emph{refs.}~\cite{Flambaum1994-fp,Lackenby2018-up,Flambaum2014-qn,Lackenby2018-ua} is underway.

\section{Appendix: Approximate Scaling of $\beta_2$}
\label{sec:A}

In this section, we verified the trend of $\beta_2\propto A^{1/3}$ established in \emph{ref.} \cite{Khriplovich2012-no} (P. 197). It is evident from Fig.~\ref{fig1} (Right-Top), that $\beta_2$ is clustered on either side of a magic number of nucleons. Therefore, the trend of $\beta_2$ as a function of $A^{1/3}$ must have peaks, and more importantly, troughs around roughly the cube root of twice the magic numbers.

\begin{figure}[h]
\centering
    \includegraphics[width=\textwidth]{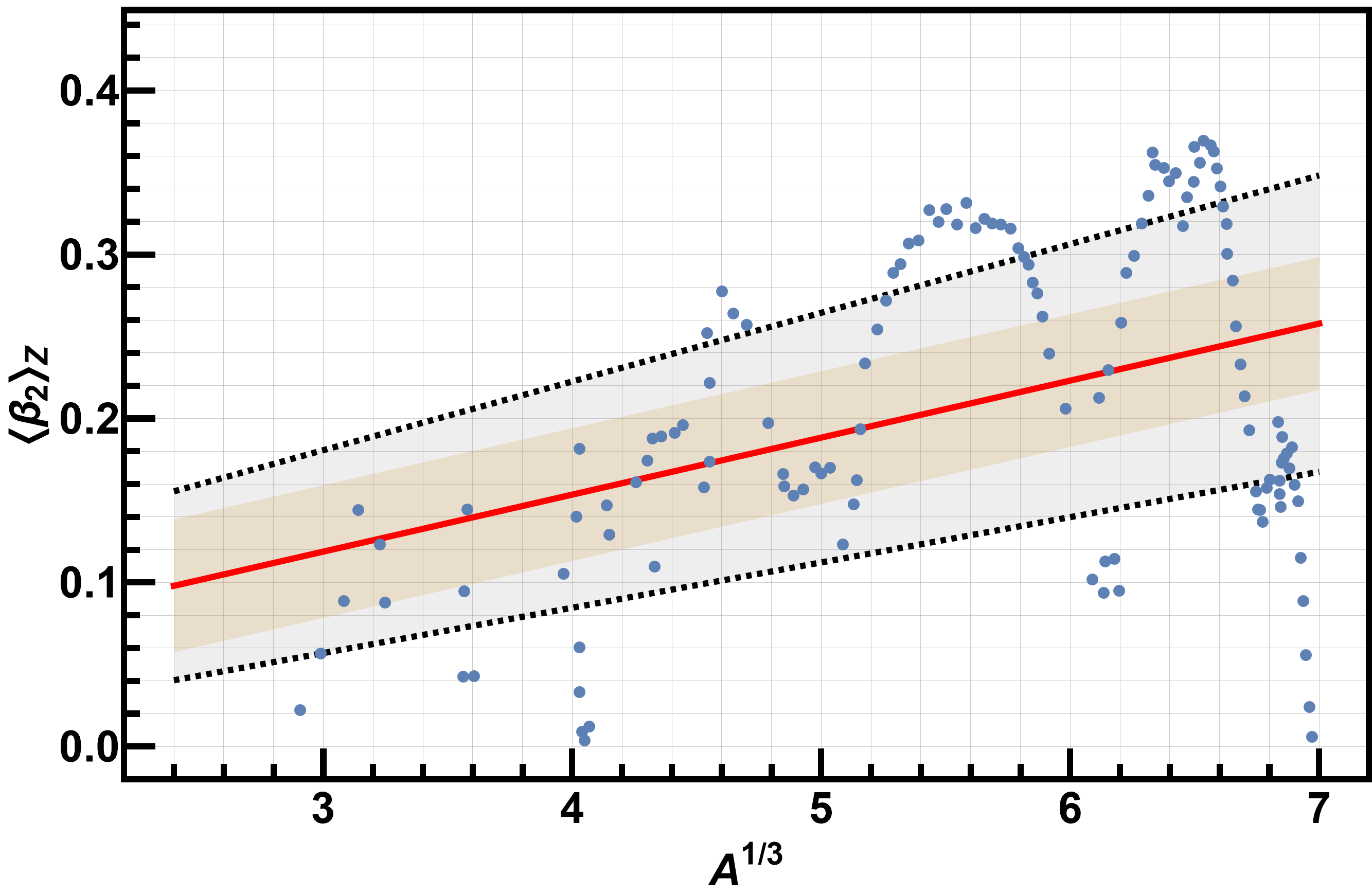}
\caption[]{Plot showing averaged maximum values of $\beta_2$ between two consecutive magic numbers of neutrons, for every fixed value for the number of protons, as a function of the cube root of their mass. The red solid line represents the best fit straight line, while the dashed lines represent the confidence interval corresponding to $68.3\%$. Darker orange region presents the confidence interval corresponding to $68.3\%$ for the best fit straight line if the uncertainty of its slope was neglected and only the uncertainty of its vertical intercept was considered.}
\label{fig2}
\end{figure}

When looking at a single element, along a horizontal line of Fig.~\ref{fig1} (Right-Top), the quadrupole deformation usually increases and then decreases as we increase the number of neutrons between two magic numbers of neutrons. For every value of $A$, there could be multiple isotopes, with varying number of protons and neutrons. In order to avoid mixing up the elements, when only considering the total number of nucleons, $A$, and in order to reduce the 3D data in Fig.~\ref{fig1} (Right-Top) to a 2D data set, we averaged the maximum values of $\beta_2$ between two consecutive magic numbers of neutrons, for every fixed value of $Z$. Such an averaged value of $\beta_2$ is indicated by $\langle\beta_2\rangle_z$ and is plotted as a function of $A^{1/3}$ in Fig.~\ref{fig2}.

We clearly observe a dip in quadrupole deformation around roughly the cube root of twice the magic numbers, $A^{1/3}\sim\{2.5,3.4,3.8,4.8,6.3\}$, representing the collective nature of $\beta_2$. We also clearly observe a general linear trend of increasing quadrupole deformation with increasing $A^{1/3}$, consistent with the trend set forth in literature \cite{Khriplovich2012-no,Lackenby2018-up,Brown2017-kh}.

\begin{acknowledgements}
One of the authors, P. M., would like to acknowledge support from Sigma Xi grants \# G2017100190747806 and \# G2019100190747806, US-Dept. of Education grant \# F-19124368008, and US-Dept. of Energy grant \#DE-SC0019768. P. M. and J. A. W. are supported by US-Dept. of Energy grant \#DE-SC0014448. We would like to thank J. Wacior for data transcription services.
\end{acknowledgements}

\section*{Conflict of interest}
The authors declare that they have no conflict of interest.

\end{document}